\documentclass[twocolumn,showpacs,preprintnumbers,amsmath,amssymb,floatfix,a4paper]{revtex4-1}  
\usepackage{epsfig}
\usepackage{dcolumn} 
\usepackage{bm}

\begin{document}
\unitlength1cm

\title{\boldmath 
Recoil-$\alpha$-fission and recoil-$\alpha$-$\alpha$-fission events
observed in the reaction $^{48}$Ca+$^{243}$Am
 \unboldmath}  

\author{
U.~Forsberg,$^1$
D.~Rudolph,$^1$
L.-L.~Andersson,$^2$
A.~Di Nitto,$^3$
Ch.E.~D\"ullmann,$^{2,3,4}$
J.M.~Gates,$^5$
P.~Golubev,$^1$
K.E.~Gregorich,$^5$
C.J.~Gross,$^6$
R.-D.~Herzberg,$^7$
F.P.~He{\ss}berger,$^{2,4}$
J.~Khuyagbaatar,$^2$
J.V.~Kratz,$^3$
K.~Rykaczewski,$^6$
L.G.~Sarmiento,$^1$
M.~Sch\"adel,$^{4,8}$
A.~Yakushev,$^4$
S.~{\AA}berg,$^1$
D.~Ackermann,$^4$
M.~Block,$^4$
H.~Brand,$^4$
B.G.~Carlsson,$^1$
D.~Cox,$^7$
X.~Derkx,$^{2,3}$
J.~Dobaczewski,$^{9}$
K.~Eberhardt,$^{2,3}$
J.~Even,$^2$
C.~Fahlander,$^1$
J.~Gerl,$^4$
E.~J\"ager,$^4$
B.~Kindler,$^4$
J.~Krier,$^4$
I.~Kojouharov,$^4$
N.~Kurz,$^4$
B.~Lommel,$^4$
A.~Mistry,$^7$
C.~Mokry,$^{2,3}$
W.~Nazarewicz,$^{6,9,10}$
H.~Nitsche,$^5$
J.P.~Omtvedt,$^{11}$
P.~Papadakis,$^7$
I.~Ragnarsson,$^1$
J.~Runke,$^4$
H.~Schaffner,$^4$
B.~Schausten,$^4$
Yue Shi,$^{6,10}$
P.~Th\"orle-Pospiech,$^{2,3}$
T.~Torres,$^4$
T. Traut,$^3$
N. Trautmann,$^3$
A.~T\"urler,$^{12}$
A.~Ward,$^7$
D.E.~Ward,$^1$
N.~Wiehl,$^{2,3}$
}

\affiliation{$^1$ Lund University, 22100 Lund, Sweden}
\affiliation{$^2$ Helmholtz Institute Mainz, 55099 Mainz, Germany} 
\affiliation{$^3$ Johannes Gutenberg-Universit\"at Mainz, 
55099 Mainz, Germany}
\affiliation{$^4$ GSI Helmholtzzentrum f\"ur Schwerionenforschung GmbH, 
64291 Darmstadt, Germany} 
\affiliation{$^5$ Lawrence Berkeley National Laboratory, Berkeley, 
California 94720, USA}
\affiliation{$^6$ Oak Ridge National Laboratory, Oak Ridge, Tennessee 37831, USA}
\affiliation{$^7$ University of Liverpool, Liverpool L69~7ZE, United Kingdom}
\affiliation{$^8$ Advanced Science Research Center, 
Japan Atomic Energy Agency, Tokai, Ibaraki 319-1195, Japan}
\affiliation{$^9$ University of Warsaw, 00681 Warsaw, Poland}
\affiliation{$^{10}$ University of Tennessee, Knoxville, Tennessee 37996, USA}
\affiliation{$^{11}$ University of Oslo, 0315 Oslo, Norway}
\affiliation{$^{12}$ Paul Scherrer Institute and University of Bern, 
5232 Villigen, Switzerland}

\date{\today}
\begin{abstract}
Products of the fusion-evaporation reaction $^{48}$Ca~+~$^{243}$Am were
studied with the TASISpec set-up at the gas-filled separator TASCA at 
the GSI Helmholtzzentrum f\"ur Schwerionenforschung. Amongst the detected 
thirty correlated $\alpha$-decay chains associated with the production
of element $Z=115$, two recoil-$\alpha$-fission and five 
recoil-$\alpha$-$\alpha$-fission events were observed. The latter are similar 
to four such events reported from experiments performed at the Dubna gas-filled
separator. Contrary to their interpretation, we propose an alternative view, namely to assign eight of 
these eleven decay chains of recoil-$\alpha$(-$\alpha$)-fission type to 
start from the $3n$-evaporation channel $^{288}$115. The other three decay 
chains remain viable candidates for the $2n$-evaporation channel $^{289}$115.
\end{abstract} 

\maketitle

In the quest for enhanced nuclear stability in the region of 
superheavy elements (SHE) -- frequently defined as those with 
$Z\geq 104$ -- the two elements flerovium (Fl, $Z=114$) and livermorium 
(Lv, $Z=116$) have recently been officially approved \cite{JWG}. The 
interpretation of data on odd-$Z$ elements, however, is more challenging, 
while the study of nuclei with odd numbers of nucleons can be especially 
rewarding: The extra hindrance and thus delay for spontaneous fission 
(SF) renders other decay modes such as $\alpha$ decay and electron capture (EC)
 even more likely. Consequently, odd-$Z$ nuclei 
potentially give rise to decay chains with more $\alpha$-decay members 
than even-$Z$ ones. Additionally, $\alpha$ decay of odd-$A$ or odd-odd 
nuclei most often proceeds into excited states in the daughter nucleus,
because unpaired nucleons typically tend to remain in the same 
single-particle orbitals as in the $\alpha$-decay parent \cite{boh75,sea90}. 
Observation of electromagnetic decays from these excited states can 
thus elucidate the low-lying nuclear structure of the daughter \cite{her08}. 
Such experimental studies have recently reached decay chains of element 
$Z=115$ \cite{rud13}, while observations of odd-$Z$ elements 
beyond Rg have been reported up to $Z=117$. See, e.g., 
Refs.~\cite{oga10,oga11,oga12,oga13,oga13b,khu14,mor12} and references therein.

The expected complex nuclear structure of odd-$Z$ nuclei of SHE can easily 
translate into complex $\alpha$-decay sequences. Interestingly, however, 
for the description of many of the hitherto published $Z\geq 113$ decay 
chains it appears to suffice with just one type of $\alpha$-decay sequence 
per isotope, be it $^{278}$113 \cite{mor12}, $^{287-289}$115 \cite{oga13}, 
or $^{293,294}$117 \cite{oga13b,khu14} (see also references therein). This 
simple but surprising picture might be due to limited statistics, combined 
with possibly comparable decay energies and half lives of different decay 
branches of a given isotope. 

In this paper, new data on recoil-$\alpha$-($\alpha$)-SF events stemming from 
element $Z=115$ are presented. A comprehensive assessment including 
corresponding data from Ref.~\cite{oga13} implies more complex decay sequences 
of $^{288,289}$115 than presented in previous reports on these isotopes
\cite{oga12,oga13,oga04}. It is demonstrated that the plain length of an 
$\alpha$-decay chain is not necessarily a good descriptor to define the 
reaction channel. 

The Universal Linear Accelerator (UNILAC) at the GSI Helm\-holtz\-zentrum 
f\"ur Schwerionenforschung, Darmstadt, Germany, provided a $^{48}$Ca$^{8+}$ 
heavy-ion beam with a typical intensity of $6\times 10^{12}$ ions per 
second, averaged over the pulsed structure of the UNILAC
(5~ms beam on and 15~ms beam off). The experiment was 
conducted at two beam energies: Beam integrals of $2.13(12)$ and 
$3.89(23)\times 10^{18}$ $^{48}$Ca ions were collected at 5.400 and 
5.462~MeV/u, respectively.

At the entrance of the recoil separator TASCA \cite{sch07,sem08,gat11}
the beam particles hit one out of four target segments, which were mounted
on a rotating target wheel \cite{jag14}. The thicknesses of these segments
averaged to 0.83(1)~mg/cm$^2$ $^{243}$Am$_2$O$_3$. The $^{243}$Am material 
originated from the Oak Ridge National Laboratory. At Mainz University, the 
$^{243}$Am was electroplated onto 2.20(5)~$\mu$m thick titanium backing foils 
\cite{run14}. The $^{48}$Ca beam first had to pass through these foils.
Estimates for the energy-loss of $^{48}$Ca ions in titanium and 
$^{243}$Am$_2$O$_3$ lead to mid-target beam energies of 242.1 
and 245.0~MeV \cite{SRIM}. Based on the Myers-Swiatecki mass
table \cite{MyS96}, these laboratory energies convert into compound nucleus 
excitation energies of $E^*=32.4$-$37.9$~MeV 
and $34.8$-$40.3$~MeV 
across the target layers.

TASCA, filled with He-gas at $p_{\rm He}=0.8$~mbar \cite{khu12}, 
was used in its so-called high-transmission mode and set to 
center ions with a magnetic rigidity of $B\rho=2.21$~Tm in the focal plane 
for the major part of the experiment \cite{for12}. The multi-coincidence 
spectroscopy set-up TASI\-Spec \cite{and10} was placed in TASCA's focal plane. 
The efficiency for transmitting element $Z=115$ fusion-evaporation residues 
through TASCA and into TASISpec amounts to 30(3)\% \cite{for12,gre13}. 

Five $32\times 32$-strip double-sided si\-licon strip detectors (DSSSD)
form the heart of TASISpec. The ions passing TASCA are
implanted in one of 1024 pixels of the downstream DSSSD, which is 
$6 \times 6$~cm$^2$ in area and 0.52~mm thick. Four additional DSSSDs of 
the same size but 1.0~mm thick are placed upstream. They are sensitive to 
charged-particle decay radiation emitted from the implanted ions into the 
backward hemisphere. Detector strips of these four DSSSDs were paired 
together electrically, i.e., they are handled effectively as 
$16\times 16$-strip DSSSDs giving rise to another $1024$ pixels. 
To detect photons coincident with charged-particle decays registered by the
DSSSDs, five large, composite germanium detectors were placed closely behind 
each of the five DSSSDs \cite{and10}.

The $96$ preamplified signals \cite{gol13} from the $n$-doped sides of the 
DSSSDs were recorded based on standard analog electronics \cite{and10}. 
The preamplified signals of the $p$-doped sides were digitized as 70-$\mu$s 
long traces by 60~MHz, 12-bit sampling ADCs \cite{FEBEX}. The signals of the
germanium detectors were handled by commercial 100-MHz, 16-bit sampling ADCs.
The data acquisition was triggered by a coincident signal from a $p$-side and 
$n$-side strip of the implantation detector. The latter limits the \mbox{energy} 
threshold of the trigger to some 400~keV deposited. This is a condition, which 
is in practice always fulfilled by 10~MeV $\alpha$-decays. Time-averaged 
(cf.~pulsed UNILAC beam structure) trigger rates were typically 100-120 events 
per second. Beam on-off status and irradiated target segment number were 
recorded as well. The data acquisition system also provided the possibility 
to send a signal to the UNILAC control system to switch off the primary 
$^{48}$Ca beam upon detection of an 8.5-11.0-MeV particle in one of the 
$n$-side strips of the implantation DSSSD during UNILAC beam-off periods. 
The beam was then chopped within $20$~$\mu$s for periods of 5-60~seconds 
\cite{rud13}.

Si- and Ge-detector calibrations were performed using various radioactive 
sources in conjunction with precision pulser signals. During the off\/line 
data analysis, the calibrations were cross-checked with known $\alpha$-decay 
as well as $\gamma$-ray energies of background radiation mainly from 
transfer reaction products reaching the TASISpec implantation detector. 
More details on the detector set-up, electronics, data storage, and data 
analysis can be found in Refs.~\cite{rud13,for14a,rud14a,rud14b,for14}.

As outlined in Ref.~\cite{rud13}, a search for time- and
position-correlated recoil-$\alpha$-$\alpha$, recoil-$\alpha$-SF, and recoil-SF
sequences was conducted using\\
$\bullet$ $\,11.5<E_{\rm rec}< 18.0$~MeV, beam on;\\
$\bullet$ $\,\,\,\,9.0<E_{\rm\alpha 1}< 12.0$~MeV, 
   $\Delta t_{\rm rec-\alpha 1}=5$~s, beam off, or\\
\hspace*{1em}$10.0<E_{\rm\alpha 1}< 12.0$~MeV, 
   $\Delta t_{\rm rec-\alpha 1}=1$~s, beam on;\\
$\bullet$ $\,\,\,\,9.0<E_{\rm\alpha 2}< 11.0$~MeV, 
   $\Delta t_{\rm\alpha 1-\alpha 2}=20$~s, beam off, or\\
\hspace*{1.25em} $9.5<E_{\rm\alpha 2}< 11.0$~MeV, 
   $\Delta t_{\rm\alpha 1-\alpha 2}=5$~s, beam on;\\
$\bullet$ $\,\,E_{\rm SF}>120$~MeV, $\Delta t_{\rm\alpha 1-SF}=30$~s, 
beam off.

During beam-off and background measurement periods, only 64 fission events 
were recorded in total. Most of these were correlated with one of the thirty 
$\alpha$-decay chains, or could be associated
with short-lived recoil-SF events of transfer reaction products such 
as $^{242m}$Am. 

\begin{figure*}[b]
\hspace*{-2mm}
\includegraphics[width=12cm]{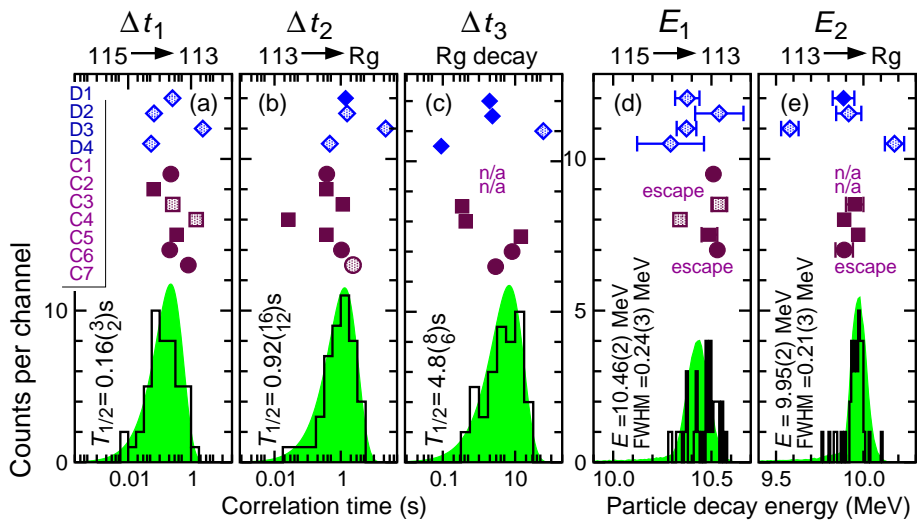}
\caption{
\label{fig1-RC}
(Color online) Correlation times [(a)-(c)] and particle decay-energy spectra 
[(d) and (e)] of decay chains observed in the reaction $^{48}$Ca+$^{243}$Am. 
The black histograms are reference spectra and relate to experimental
data from the 53 chains associated with the $3n$ evaporation channel 
$^{288}$115 in Refs.~\cite{rud13,oga13,rud14a}. 
The shaded areas in panels (a)-(c) are the corresponding 
time distributions. The shaded areas in panels (d) and (e) are 
the energy distributions derived from Monte Carlo simulations as outlined
in Refs.~\cite{rud13,rud14b,sar14}. The data points of the eleven observed 
recoil-$\alpha$-($\alpha$-)fission events are provided on top of each
spectrum, labeled with an identifier according to Table~\ref{tab:chains}.
Filled symbols indicate measurement during beam-off periods. Diamonds 
refer to data from Ref.~\cite{oga13}, and squares and circles refer 
to the present data observed at 242.1 and 245.0~MeV mid-target beam energy,
respectively. `n/a' means `not applicable', `escape' denotes events with 
low-energy detection solely in the implantation DSSSD.}
\end{figure*}

The thirty identified chains associated with the production of element 115
contain 23 five-$\alpha$ long chains. The spectroscopic results on these
have been communicated in Ref.~\cite{rud13}, further described in 
Refs.~\cite{rud14a,rud14b}, and are to be detailed in a forthcoming 
publication \cite{for14}. One of those 23 long chains was assigned to 
originate from the isotope $^{287}$115, and 22 from $^{288}$115 \cite{rud13}. 
The combined data of these 22 decay chains and 31 corresponding ones from 
Ref.~\cite{oga13} are shown at the bottom of Fig.~\ref{fig1-RC}. The black 
histograms in Fig.~\ref{fig1-RC} are the experimental spectra. 
In panels (a)-(c), the number of correlation times available 
to derive the half-life, $T_{1/2}$, of a given decay step is 47, 46, and 47, 
respectively. The expected time distributions for the corresponding
half-lives are indicated by the shaded areas.
In Figs.~\ref{fig1-RC}(d) and (e) the shaded areas relate to Geant4 
Monte-Carlo simulations \cite{sar12,sar14}, which are based on decay schemes 
suggested in Refs.~\cite{rud13,rud14a,rud14b,sar14}. In short, one or several 
$\alpha$ decays populate excited states in the daughter nucleus. The
excited states decay primarily via internal conversion. 
Energy summing of a detected $\alpha$ particle and one or more registered 
conversion or Auger electrons readily explains the relatively broad energy 
distributions observed.

To assess whether distributions of experimental correlation times are 
compatible with the assumption that each step can be described by one single 
half-life, a relatively ``new test for random events of an exponential 
distribution'' \cite{sch00} was applied to the 53 chains. This is done 
({\it i}) for the full data set, and ({\it ii}) subdivided into six data sets 
corresponding to the six different $^{48}$Ca beam energies 
exploited in Refs.~\cite{rud13,oga13}. The result of this
test can be found in Table~IV and Fig.~4 of the Supplemental Material (SM) 
\cite{SuppMat}. There is no clear hint towards the 
need of assuming the decay of more than one radioactive species for any of the 
decay steps of these 53 chains, thereby concurring with previous 
interpretations \cite{oga13,rud14a,rud14b}. Therefore, and this is the most 
relevant result in the context of the present work, they serve as reference
for the $3n$ evaporation channel $^{288}$115.

\begin{figure*}[p]
\includegraphics[width=18cm]{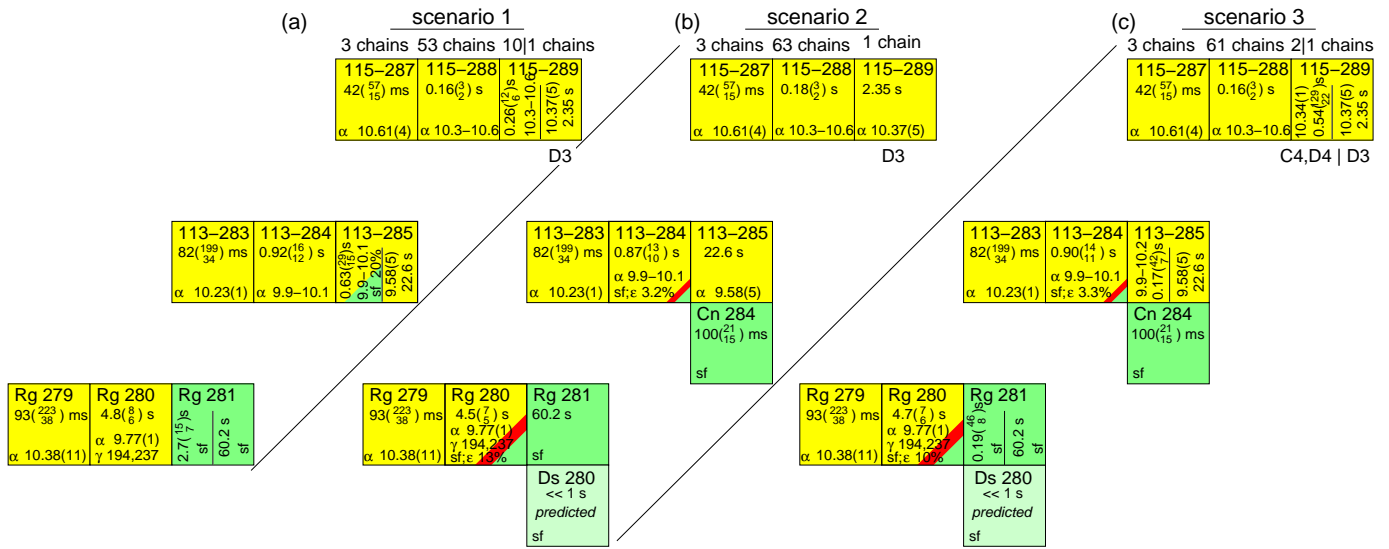}
\caption{
\label{fig2-RC}
(Color online) Three different decay scenarios of directly produced
isotopes of element $Z=115$ down to Rg, using previously published 
\protect\cite{rud13,oga13} and present data. Half-lives, $T_{1/2}$, are 
provided with uncertainties. Particle-decay energies or ranges of decay 
energies are given in MeV, $\gamma$-ray energies in keV.  
See text for detailed discussions.
}
\end{figure*}

The focus lies now on the interpretation of two recoil-$\alpha$-SF 
(denoted C1,C2) and five recoil-$\alpha$-$\alpha$-SF (C3-C7) events 
from the present work, and four similar recoil-$\alpha$-$\alpha$-SF 
chains published by Oganessian {\it et al.} (D1-D4) \cite{oga13}.
Three decay scenarios are illustrated in Fig.~\ref{fig2-RC} and discussed
below: In `scenario 1', all eleven recoil-$\alpha$(-$\alpha$)-SF chains
are associated with $^{289}$115, with chain D3 forming a distinctively 
different sequence. In `scenario 2', all recoil-$\alpha$-($\alpha$)-SF 
chains but D3 are considered to originate from $^{288}$115. For 'scenario 3',
two chains, C4 and D4, are moved from $^{288}$115 to $^{289}$115, forming
a decay sequence parallel to chain D3. On purpose, we refrain from using 
decay data suggested to originate from $^{293}$117 
\cite{oga10,oga11b,oga12,oga13b}. Presence of links between decay 
chains associated with elements 115 and 117 are neither questioned nor 
subject of the present work.

The experimental results of the eleven recoil-$\alpha$(-$\alpha$)-SF 
chains are summarized in Table~\ref{tab:chains}. 
The number of chains of a given type expected
to arise from random background in the whole implantation DSSSD, 
$N_{\rm random}$, is calculated according to Ref.~\cite{sch84}. Here, 
escape events relate to energies up to 4~MeV in a DSSSD pixel, 
while 9-11~MeV is set for full energy and reconstructed events. The time 
periods used in the calculations are 2~s, 10~s, and 50~s for decay steps one, 
two, and three. These correspond to roughly ten times the respective 
average half-lives of the scenarios (cf.~Fig.~\ref{fig1-RC}). Count rates per pixel are determined
in the two pre-defined energy ranges separately for beam-on and beam-off 
periods. Essentially, the non-random origins of TASISpec chains C1-C7 are 
defined by the overall small number of fission events observed during 
beam-off periods, in combination with the rather short time periods, 
$\Delta t$, between the detected recoil implantation and subsequent 
fission events. 

The individual correlation times and decay energies of the eleven 
recoil-$\alpha$-($\alpha$)-SF chains are also provided in 
Fig.~\ref{fig1-RC}(a)-(e), labeled with C1-C7 and D1-D4, respectively. 
At first glance, there seems to be very little if any difference in the 
distribution of these data points compared with the suggested reference 
distributions of the 53 five-$\alpha$ long chains associated with $^{288}$115
(see above). Interestingly, none of the previous publications on decay chains 
associated with elements 115 and 117 (see, e.g., 
Refs.~\cite{oga11,oga13,oga11b,oga12,oga13b} 
dwells on this obvious similarity. Instead, it has essentially been 
argued that the length of a decay chain is a sufficient descriptor of
its origin: For example, all five-$\alpha$ long chains observed at low
excitation energies of the compound nucleus $^{291}$115, $E^*<37$~MeV, 
were associated with the decay of the $3n$ evaporation 
channel $^{288}$115, whilst chains D1-D4 were interpreted to originate 
from the $2n$ evaporation channel $^{289}$115 \cite{oga13}. 

Columns 2 (`$3n$') and 3 (C1-C7, D1-D4) of Table~\ref{tab:times} provide the 
half-life analyses of correlation times. This very simple 
decay scenario, essentially based on the observed length of a decay chain, 
implies a ratio of maximum production cross-sections 
$R_\sigma=\sigma(2n)/\sigma(3n)\approx 0.5$ 
(cf.~Ref.~\cite{oga13} and Table~\ref{tab:xsec}). 
Such high relative or absolute yields of the production of the $2n$ reaction channel $^{289}$115 are
not necessarily consistent with expectations from nuclear reaction theory 
\cite{zag04,siw12}. Systematic studies of fusion-evaporation reactions 
leading to heavy or superheavy nuclei \cite{rei92} also prefer 
$R_\sigma\lesssim$~10\%. 
Secondly, a closer look at Fig.~\ref{fig1-RC} suggests that chain D3 
is neither compatible with the $3n$ reference values nor the average
of the remaining ten recoil-$\alpha$-($\alpha$)-SF events 
(cf.~column 4 of Table~\ref{tab:times}): all three decay times of 
D3 reach significantly beyond the upper limit of the respective reference
time distribution. Its energy $E_2$ is also significantly lower than that of
all other chains. While a single extreme value would not pose a statistical 
problem, the fact that in chain D3 four out of five observables differ 
significantly from the expectations does. 

This observation is turned into a more robust figure-of-merit (FoM). It provides a 
measure for the congruence of the time sequence of 
the first decay steps of a single (element 115) chain with respect to the 
average of a certain 'reference' ensemble of (element 115) chains. For each of the decay steps, $i=1,2,3$, 
 the value $P_i$ is the reference probability density function 
for correlation times in logarithmic-sized bins (see Ref.~\cite{sch00})
evaluated for the measured correlation time. Thereafter, it is normalised with the constant $e$ 
such that the values range from 0 to 1. Figure~\ref{fig1-RC} shows the form of the
 functions, here normalised to the amount of data. To account for the uncertainties in
 the reference half-lives, $P_i$ is given the value 1 whenever the correlation time is within
 the confidence limits. If it is above (below) the confidence interval, the upper (lower)
 limit is used. To be able to compare chains with differing numbers, $n$, of decay steps, 
the FoM=$\langle P_t\rangle =^n$\hspace*{-1mm}$\sqrt{\prod P_i}$ is calculated 
as the geometric average. Deviations in decay energy only serve as supportive argument, since the 
comprehensive data of element 115 decay chains rather suggests a range of 
energies than distinct peaks for decay steps $115\rightarrow 113$ and 
$113\rightarrow$~Rg \cite{rud13,rud14b}.

The rightmost columns in Table~\ref{tab:chains} list $\langle P_t\rangle$
for the three decay scenarios proposed in Fig.~\ref{fig2-RC}. 
For more detailed statistical assessments see Tables~SM~VI-X. 
In short, $\langle P_t\rangle\gtrsim 50\%$ appears to 
be a solid {\em indicator} for congruence. 
In turn, $\langle P_t\rangle\lesssim 25\%$ 
{\em signals} a decay chain different from the reference.

`Scenario 1', cf.~Fig.~\ref{fig2-RC}(a):\\ 
Outstandingly low values of $\langle P_t\rangle$(D3)=$0.10\%$, $18\%$, and 
$0.074\%$ are obtained for chain D3 when cross-checked versus the 
53 five-$\alpha$-long chains, the average of all eleven, or the remaining 
ten recoil-$\alpha$-($\alpha$)-SF chains, respectively 
(cf.~Tables~\ref{tab:chains} and SM~VI-VIII). The conclusion is that 
chain D3 is not compatible with any subset of the other element 115 decay 
chains. D3 thus represents a single but distinct decay sequence starting from 
a given state of a certain isotope of element 115. Note that it is 
{\em not} the intention to question the existence of this chain as such. 
In fact, according to our interpretations, D3 remains a valid candidate to be 
attributed to $^{289}$115 (see below). 

Note also that the decay times of D3 are
similar to the chain $^{289}$Fl$\rightarrow ^{285}$Cn$\rightarrow ^{281}$Ds 
\cite{oga11,gat11,due10}.
This decay sequence can in principle be entered by EC
decay of $^{289}$115 or $^{285}$113. However, both $\alpha$-decay energies 
measured for D3 differ distinctively from the ones expected for the 
mentioned even-$Z$ chain. Therefore, this explanation for D3 is disregarded.

`Scenario 2', cf.~Fig.~\ref{fig2-RC}(b):\\ 
In decay `scenario 2', all the other ten recoil-$\alpha$-($\alpha$)-SF 
chains -- C1-C7, D1, D2, and D4 -- 
are considered to start from the isotope $^{288}$115. 
Comparing columns 2 and 4 in Table~\ref{tab:times} and inspecting 
Fig.~\ref{fig2-RC}(a), the correlation time analysis indicates essentially 
compatible half-lives for all three decay steps. There is no significant 
FoM difference between the congruence test of these ten chains among
themselves ($\langle$FoM$\rangle=64\%$) and the values obtained when 
they are associated with decays of $^{288}$115 ($\langle$FoM$\rangle=65\%$). 
There is, however, a clear improvement compared with 
plain averages of all eleven recoil-$\alpha$-($\alpha$)-SF 
chains ($\langle$FoM$\rangle=52\%$, cf.~Tables SM~VII-IX).

The assignment of the ten chains to $^{288}$115 implies, at first sight, 
SF branching ratios of $b_{\rm SF}=2/63=3.2$\% for $^{284}$113 
and $b_{\rm SF}=8/61=13$\% for $^{280}$Rg, respectively. It is important to 
note, however, that none of the experiments has been sensitive to
EC decay. Hence, other options are EC decay branches 
of $^{284}$113 or $^{280}$Rg into 
even-even $^{284}$Cn or $^{280}$Ds. The latter are either known 
(see, e.g., Refs.~\cite{oga11,gat11,due10}) or expected (see, e.g., 
Refs.~\cite{smo95,war12,sta13}) to decay with $T_{1/2}$(SF)$\ll 1$~s. 
In this scenario, partial SF or EC half-lives amount to 
about 30~s for both $^{284}$113 and $^{280}$Rg. This suggests SF 
hindrance factors of 300 relative to $^{284}$Cn. This value is rather 
low for odd-odd nuclei, because hindrance is expected from each of
the two unpaired nucleons. Already for one unpaired nucleon,
hindrance factors are typically above 1000 \cite{hof95,hes14}.
Thus, we rather suggest EC preceding 
SF of the respective even-even daughter. The observation of SF or 
EC branches in this region of the nuclear chart is consistent with 
recent theoretical estimates \cite{kar12,zag13} (see especially 
Figs.~4 and 6 in Ref.~\cite{kar12}). For the half-life analysis in 
Table~\ref{tab:times} and Table SM~V, SF of $^{284}$113 and $^{280}$Rg has 
been assumed, i.e., the short finite lifetime of a potential EC daughter 
has not been considered.

`Scenario 3', cf.~Fig.~\ref{fig2-RC}(c):\\ 
Further examinations of Fig.~\ref{fig1-RC} and Tables SM~VII-IX reveal
that C4 and D4 may have a common origin, which is different from that of 
all other chains: $\langle P_t\rangle<25\%$ 
for both C4 and D4 in all hitherto discussed combinations, and in 
particular $E_{2}$ of chain D4 lies outside the typical window for 
$\alpha$-decay energies of $^{284}$113. The rather short correlation 
time of the observed fission event of chain D4 is also striking.
Columns 5 and 6 of Table~\ref{tab:times} provide the half-lives resulting 
from the correlation-time analysis of the remaining eight chains -- 
C1-C3, C5-C7, D1-D2 -- and C4 combined with D4, respectively. 

The numbers in Table~\ref{tab:times} indicate that there is no obvious 
reason {\em not} to 
assign the former eight chains with either SF or EC decay branches of 
decay chains starting with $^{288}$115. Chains C4 and D4 {\em may}
form a second decay sequence starting from $^{289}$115, parallel to chain D3. 
It is consistent with -- but not conditioned by -- the detection of C4 
at the lower beam energy of the present TASISpec experiment. 
Despite significant differences of the first decay steps of C4 
and D4 we {\em propose} them to stem from the same level. Combining 
these two chains is the simplest approach within the available statistics 
($\langle$FoM$\rangle=74\%$, cf.~Table SM~X). A related
scenario could include also chain D1 into this subset, in this case based on a 
selection according to decay energy of the first decay step. 

Table~\ref{tab:xsec} provides measured cross sections for the three decay 
scenarios shown in Fig.~\ref{fig2-RC}. For `scenario 1' the numbers would
be consistent with Fig.~4 of Ref.~\cite{oga13}, but it implies too high values 
for the $2n$ channel (see above). In the preferred `scenario 3', the
cross-section of the $2n$ reaction channel $^{289}$115 amounts to 
$\approx 1$~pb at low excitation energies $E^*\approx 34$~MeV. This 
corresponds to $\approx 10$\% of the maximum cross-section of the 
$3n$-channel around $E^*=37$~MeV and is in line with theoretical expectations.

\begin{figure}
\begin{center}
\includegraphics[width=7cm]{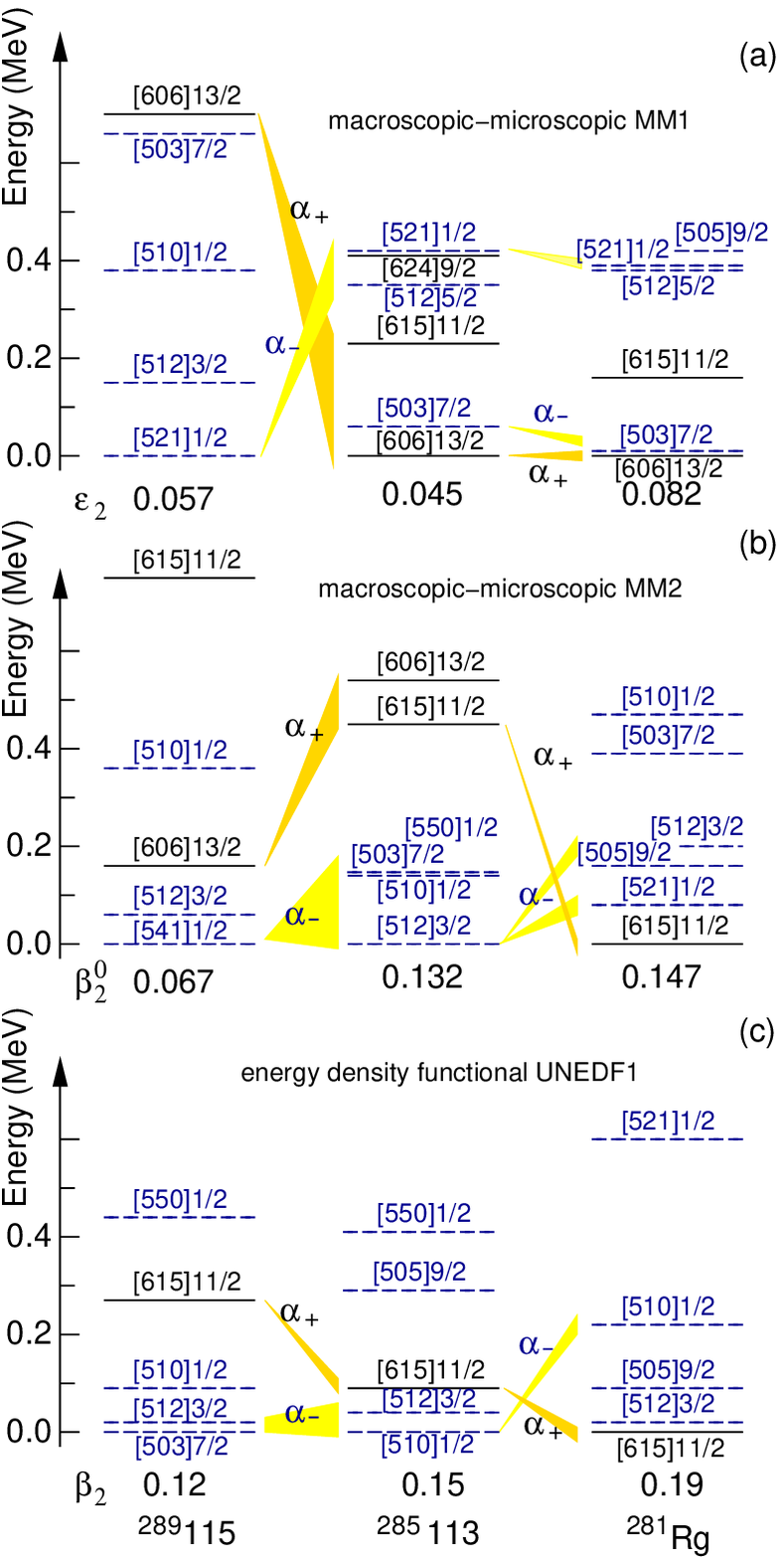}
\end{center}
\vspace*{-5mm}
\caption{
\label{fig3-RC}
(Color online) Nuclear structure predictions of low-lying
states in $^{289}$115, $^{285}$113, and $^{281}$Rg. Panels (a) and (b) are 
based on two different macroscopic-microscopic approaches, MM1 \cite{car06,nil69} and MM2 
\cite{par04}. Panel (c) originates from the self-consistent
Skyrme energy density functional UNEDF1 \cite{kor12,yue14}. 
The results from UNEDF1$^{\rm SO}$ are similar to MM2 \cite{yue14}. Proton 
single-particle states are labeled with their asymptotic Nilsson quantum 
numbers $[Nn_z\Lambda]\Omega$. Full (black) lines represent positive-parity 
states, dashed (blue) lines negative-parity states. Selected
$\alpha$-decay sequences among high-$\Omega$ positive-parity ($\alpha_+$) 
and low-$\Omega$ negative-parity ($\alpha_-$) states are indicated by shaded areas.
$\epsilon_2$, $\beta^0_2$, $\beta_2$, respectively, denote predicted 
quadrupole deformations.}
\end{figure}

Figure~\ref{fig3-RC} provides proton single-particle energies predicted
by macroscopic-microscopic model parametrisations 
\cite{car06,nil69,par04} and a Skyrme energy density functional
\cite{cwi99,kor12,yue14}. Independent of the model or the 
parametrisation, the nuclear shape is predicted to change from 
near-sphericity 
towards prolate deformation 
along the decay chain 
$^{289}$115$\rightarrow^{285}$113 $\rightarrow^{281}$Rg. 
Most interestingly, however, all models suggest the same decay {\em pattern},
namely two independent $\alpha$-decay chains: Exemplified by
Fig.~\ref{fig3-RC}(b), once $^{289}$115 is created as final fusion-evaporation 
product, excited states will decay by electromagnetic radiation into 
{\em either} a high-$\Omega$ positive-parity state (here: $[606]13/2$) 
{\em or} a low-$\Omega$ negative-parity state (here: $[541]1/2$). The
two families (high-$\Omega$ and low-$\Omega$) of Nilsson orbitals remain separate for the daughter $^{285}$113 and 
grand-daughter $^{281}$Rg, giving rise to two parallel $\alpha$-decay 
sequences. The detailed predictions are highly model dependent and it is currently not possible to say which predicted decay sequence can be associated with the observed ones. 

In summary, seven new recoil-$\alpha$-($\alpha$)-SF events were observed
following the fusion-evaporation reaction $^{48}$Ca+$^{243}$Am. An assessment
of these seven together with four previously published \cite{oga13} decay 
chains suggests a revised decay scenario of isotopes of element 115: Eight of 
these eleven decay chains are interpreted to originate from $^{288}$115. They
represent either SF or EC decay branches of $^{284}$113
and $^{280}$Rg. The remaining three chains can account for two separate decay
sequences of the isotope $^{289}$115. Two parallel sequences are readily 
explained by a set of nuclear structure models. Clearly, more high-quality 
spectroscopic data near the barrier of the 
$^{48}$Ca+$^{243}$Am reaction is needed to verify {\em any} 
\emph{proposed} decay scenario of $^{288,289}$115. This is also considered necessary 
to establish statistically firm links towards $^{293}$117. A comprehensive 
discussion of this subject is beyond the scope of the present paper. 

The authors would like to thank the ion-source and the accelerator 
staff at GSI. This work is supported by the European Community FP7 -- 
Capacities ENSAR No.~262010, the Royal Physiographic Society in Lund,
the Euroball Owners Committee, 
the Swedish Research Council, the German BMBF, the Office of Nuclear Physics,
U.S. Department of Energy, and the UK Science and Technology 
Facilities Council.

\begin{table*}[p]
\tabcolsep5pt
\caption{
\label{tab:chains}
Mid-target laboratory-frame beam energies, energies of the implanted 
recoils $E_{\rm rec}$, decay energies $E_1$, $E_2$ and $E_3$,
together with the associated correlation times of recoil-$\alpha$-SF and
recoil-$\alpha$-$\alpha$-SF decay chains observed in the 
$^{48}$Ca+$^{243}$Am reaction. Entries in bold were recorded during beam-off 
periods. The number of $\gamma$ rays detected in prompt coincidence with 
SF events, $N_\gamma$(SF), are also specified. $N_{\rm random}$ corresponds
to the number of chains of a given type expected to arise from random background 
\protect\cite{sch84}.
$\langle P_t\rangle$ represents a figure-of-merit for the congruence of a 
single chain with a proposed decay scenario. See text, Fig.~\ref{fig2-RC}, and
Supplemental Material for details. The decay characteristics
of the four chains listed in Table III in Ref.~\protect\cite{oga13} are 
included for completeness, denoted D1-D4.}
\begin{tabular}{ccccccccccc}
\hline
& \\[-8pt]
No.& $\langle E_{\rm lab}\rangle$ & $E_{\rm rec}$ (MeV) &$E_1$ (MeV)&$E_2$ (MeV)&$E_3$ (MeV) & $N_\gamma$(SF) &
   $N_{\rm random}$ &\multicolumn{3}{c}{$\langle P_t\rangle$ for scenario$^d$} \\
   &  (MeV)     & pixel (x,y)     &$\Delta t_1$ (s)  &$\Delta t_2$ (s)  &$\Delta t_3$ (s) & & & 1 & 2 & 3 \\
\hline
& \\[-8pt]
C1 & 245.0 & 12.3 & {\bf 10.51(1)} &    {\bf 242}$^a$   & & {\bf 6} &$< 2\cdot 10^{-5}$ & 91 & 81 & 80\\
   &       & 268 (8,12) & {\bf 0.227}    &    {\bf 0.378}     \\
C2 & 242.1 & 16.2 & {\bf 1.45(1)}$^b$ & {\bf 211}       & & {\bf\boldmath $> 4$\unboldmath} & $< 6\cdot 10^{-2}$ & 64 & 61 & 63 \\
   &       & 425 (13,9) & {\bf 0.0645}      & {\bf 0.366}     \\
C3 & 242.1 & 13.9 & 10.54(4)$^c$ & {\bf 9.95(5)}$^c$ & {\bf 196} & {\bf 8} & $< 2\cdot 10^{-6}$ & 49 & 53 & 53 \\
   &       & 681 (21,9) & 0.261        & {\bf 1.15}        & {\bf 0.343} \\
C4 & 242.1 & 14.5 & 10.34(1)     & {\bf 9.89(1)}     & {\bf 218}$^a$ & {\bf\boldmath $> 5$\unboldmath}
& $< 2\cdot 10^{-6}$ & 19 & 11 & 74 \\
   &       & 344 (10,24)& 1.46         & {\bf 0.0262}      & {\bf 0.432} \\
C5 & 242.1 & 13.8 &{\bf 10.49(4)}$^c$ & {\bf 9.97(1)}& {\bf 135} & {\bf 9} & $< 3\cdot 10^{-9}$& 80 & 79 & 79 \\
   &       & 554 (17,10)&{\bf 0.345}        & {\bf 0.369}  & {\bf 14.4} \\
C6 & 245.0 & 14.5 & {\bf 10.53(1)}& {\bf 9.89(5)}$^c$ & {\bf 230}$^a$& {\bf 9}& $< 3\cdot 10^{-9}$& 97 & 100 & 100 \\
   &       & 205 (6,13) & {\bf 0.210}   & {\bf 1.05}        & {\bf 8.27}\\
C7 & 245.0 & 11.9 & {\bf 0.541(3)}$^b$& 3.12(1)$^b$   & {\bf 230}$^a$ & {\bf \boldmath $> 4$\unboldmath} 
& $< 1\cdot 10^{-1}$ & 81 & 71 & 67 \\ 
   &       & 128 (4,0)  & {\bf 0.815}       & 2.33        & {\bf 2.89}\\
\hline
D1 & 240.5 & 11.38& 10.377(62) & {\bf 9.886(62)} & {\bf 215.7} &&& 82 & 87 & 87 \\
   &       &      & 0.2562     & {\bf 1.4027}    & {\bf 1.9775}\\
D2 & 241.0 & 15.18& 10.540(123)$^c$& 9.916(72)   & {\bf 214.9}$^a$ &&& 67 & 76 & 78 \\
   &       &      & 0.0661     & 1.5500          & {\bf 2.3638}\\
D3 & 241.0 & 9.04 & 10.373(50) & 9.579(50)       & 141.1 &&& n/a & n/a & n/a\\
   &       &      & 2.3507     & 22.5822         & 60.1855\\
D4 & 241.0 & 13.35& 10.292(170)$^c$& 10.178(55)  & {\bf 182.2}$^a$&&& 23 & 25 & 63 \\
   &       &      & 0.0536     & 0.4671          & {\bf 0.0908}\\
\hline
\end{tabular}

\footnotemark[1]{Fission event registered by both implantation and box detector.}\\
\footnotemark[2]{Escaped $\alpha$ particle registered solely by the implantation detector.}\\
\footnotemark[3]{Reconstructed energy of an $\alpha$ particle registered by both implantation and box detector.}\\
\footnotemark[4]{For details see Tables~VII, IX and X of the Supplemental Material.}\\
\end{table*}

\begin{table}
\tabcolsep3.5pt
\caption{
\label{tab:times}
Half-lives, $T_{1/2}$,  derived from the correlation times of decays of 
isotopes of $Z=115$, $Z=113$ and Rg. The column labeled `$3n$' 
relates to the reference decay chains associated with the $3n$ reaction 
channel in Refs.~\protect\cite{oga13,rud13,rud14a}. The other columns 
describe different scenarios (cf.~Fig.~\protect\ref{fig2-RC}) of the 
decay data from recoil-$\alpha$-SF and recoil-$\alpha$-$\alpha$-SF events.
}
\begin{tabular}{cccccc}
\hline
& \\[-8pt]
data    & $3n$ & C1-C7 & C1-C7  & C1-C3 & C4,D4 \\
selection&\protect\cite{oga13,rud13}
& D1-D4 &D1,D2,D4& C5-C7 &       \\
&&&& D1,D2\\
\hline
& \\[-8pt]
& $T_{1/2}$ (s) &$T_{1/2}$ (s) &$T_{1/2}$ (s) &$T_{1/2}$ (s) &$T_{1/2}$ (s) \\
\hline
& \\[-8pt]
$Z=115$ & 
$0.16(^3_2)$      & $0.39(^{17}_9)$ & $0.26(^{12}_6)$ & 
$0.19(^{11}_5)$  & $0.54(^{129}_{22})$\\
& \\[-8pt]
$Z=113$ &
$0.92(^{16}_{12})$& $2.0(^9_5)$ & $0.63(^{29}_{15})$ &
$0.75(^{41}_{20})$  &$0.17(^{42}_{7})$\\
& \\[-8pt]
Rg       &
 $4.8(^8_6)$      & $7.0(^{35}_{18})$ & $2.7(^{15}_7)$ &
 $3.5(^{24}_{10})$  & $0.19(^{46}_8)$\\  
\hline
\end{tabular}
\end{table}

\begin{table*}
\caption{
\label{tab:xsec}
Number of chains and production cross sections, $\sigma_{\rm prod}$, 
of $^{287-289}$115 derived for three decay scenarios 
(cf.~Fig.~\ref{fig2-RC}) from the thirty decay chains observed
in the TASISpec experiment (chains C1-C7 and
23 five-$\alpha$ long chains \cite{rud13}). The observation of a single decay chain relates to a production cross section of $\sigma=0.93(12)$ and $0.51(7)$~pb for mid-target beam energies 242.1 and 245.0~MeV, respectively. The systematic 
uncertainty accounts for uncertainties in beam integrals, target thickness,
transport efficiency, and identification probability.
Standard deviations of systematic uncertainties are given together 
with statistical uncertainties using a 68\% 
confidence level \protect\cite{sch84}.}
\begin{tabular}{ccc|cc|cc|cc}
\hline
& \\[-8pt]
&&&\multicolumn{2}{c}{\hspace*{13mm}scenario 1\hspace*{13mm}}&
   \multicolumn{2}{c}{\hspace*{13mm}scenario 2\hspace*{13mm}}&
   \multicolumn{2}{c}{\hspace*{13mm}scenario 3\hspace*{13mm}} \\
$\langle E_{\rm lab}\rangle$&$E^*$ (MeV) &
reaction&no. of& $\sigma_{\rm prod}$& no. of& $\sigma_{\rm prod}$&
no. of& $\sigma_{\rm prod}$\\ 
(MeV) & \protect\cite{MyS96}  &channel &chains& (pb)           & 
chains& (pb)           & chains& (pb)           \\
\hline
& \\[-8pt]
242.1 & 32.4-37.9 &$2n$ & 4     & $3.7\pm 0.5^{+2.8}_{-1.9}$& 
                          0     & $<1.9$ &
                          1     & $0.93\pm 0.12^{+2.14}_{-0.77}$ \\
& \\[-8pt]
      &           &$3n$ & 8     & $7.5\pm 1.0^{+3.6}_{-2.6}$&
                          12    & $11.2\pm 1.4^{+4.2}_{-3.2}$&
                          11    & $10.3\pm 1.3^{+4.0}_{-3.1}$ \\
& \\[-8pt]
      &           &$4n$ & 0     & $<1.9$ &
                          0     & $<1.9$ &
                          0     & $<1.9$ \\
& \\[-8pt]
245.0 & 34.8-40.3 &$2n$ & 3     & $1.5\pm 0.2^{+1.4}_{-0.9}$& 
                          0     & $<1.1$ &
                          0     & $<1.1$\\
& \\[-8pt]
      &           &$3n$ & 14    & $7.2\pm 0.9^{+2.4}_{-1.9}$& 
                          17    & $8.7\pm 1.1^{+2.6}_{-2.1}$&      
                          17    & $8.7\pm 1.1^{+2.6}_{-2.1}$ \\
& \\[-8pt]
      &           &$4n$ & 1     & $0.51\pm 0.07^{+1.17}_{-0.42}$&  
                          1     & $0.51\pm 0.07^{+1.17}_{-0.42}$&  
                          1     & $0.51\pm 0.07^{+1.17}_{-0.42}$  \\
& \\[-8pt]
\hline
\end{tabular}
\end{table*}

\end{document}


\unitlength1cm

\title{\boldmath 
Recoil-$\alpha$-fission and recoil-$\alpha$-$\alpha$-fission events
observed in the reaction $^{48}$Ca+$^{243}$Am
 \unboldmath}  

\author{
U.~Forsberg,$^1$
D.~Rudolph,$^1$
L.-L.~Andersson,$^2$
A.~Di Nitto,$^3$
Ch.E.~D\"ullmann,$^{2,3,4}$
J.M.~Gates,$^5$
P.~Golubev,$^1$
K.E.~Gregorich,$^5$
C.J.~Gross,$^6$
R.-D.~Herzberg,$^7$
F.P.~He{\ss}berger,$^{2,4}$
J.~Khuyagbaatar,$^2$
J.V.~Kratz,$^3$
K.~Rykaczewski,$^6$
L.G.~Sarmiento,$^1$
M.~Sch\"adel,$^{4,8}$
A.~Yakushev,$^4$
S.~{\AA}berg,$^1$
D.~Ackermann,$^4$
M.~Block,$^4$
H.~Brand,$^4$
B.G.~Carlsson,$^1$
D.~Cox,$^7$
X.~Derkx,$^{2,3}$
J.~Dobaczewski,$^{9}$
K.~Eberhardt,$^{2,3}$
J.~Even,$^2$
C.~Fahlander,$^1$
J.~Gerl,$^4$
E.~J\"ager,$^4$
B.~Kindler,$^4$
J.~Krier,$^4$
I.~Kojouharov,$^4$
N.~Kurz,$^4$
B.~Lommel,$^4$
A.~Mistry,$^7$
C.~Mokry,$^{2,3}$
W.~Nazarewicz,$^{6,9,10}$
H.~Nitsche,$^5$
J.P.~Omtvedt,$^{11}$
P.~Papadakis,$^7$
I.~Ragnarsson,$^1$
J.~Runke,$^4$
H.~Schaffner,$^4$
B.~Schausten,$^4$
Yue Shi,$^{6,10}$
P.~Th\"orle-Pospiech,$^{2,3}$
T.~Torres,$^4$
T. Traut,$^3$
N. Trautmann,$^3$
A.~T\"urler,$^{12}$
A.~Ward,$^7$
D.E.~Ward,$^1$
N.~Wiehl,$^{2,3}$
}

\affiliation{$^1$ Lund University, 22100 Lund, Sweden}
\affiliation{$^2$ Helmholtz Institute Mainz, 55099 Mainz, Germany} 
\affiliation{$^3$ Johannes Gutenberg-Universit\"at Mainz, 
55099 Mainz, Germany}
\affiliation{$^4$ GSI Helmholtzzentrum f\"ur Schwerionenforschung GmbH, 
64291 Darmstadt, Germany} 
\affiliation{$^5$ Lawrence Berkeley National Laboratory, Berkeley, 
California 94720, USA}
\affiliation{$^6$ Oak Ridge National Laboratory, Oak Ridge, Tennessee 37831, USA}
\affiliation{$^7$ University of Liverpool, Liverpool L69~7ZE, United Kingdom}
\affiliation{$^8$ Advanced Science Research Center, 
Japan Atomic Energy Agency, Tokai, Ibaraki 319-1195, Japan}
\affiliation{$^9$ University of Warsaw, 00681 Warsaw, Poland}
\affiliation{$^{10}$ University of Tennessee, Knoxville, Tennessee 37996, USA}
\affiliation{$^{11}$ University of Oslo, 0315 Oslo, Norway}
\affiliation{$^{12}$ Paul Scherrer Institute and University of Bern, 
5232 Villigen, Switzerland}

\date{\today}

\maketitle

\addtocounter{table}{3}
\addtocounter{figure}{3}

\begin{figure}[ht]
\hspace*{-2mm}
\includegraphics[width=9cm]{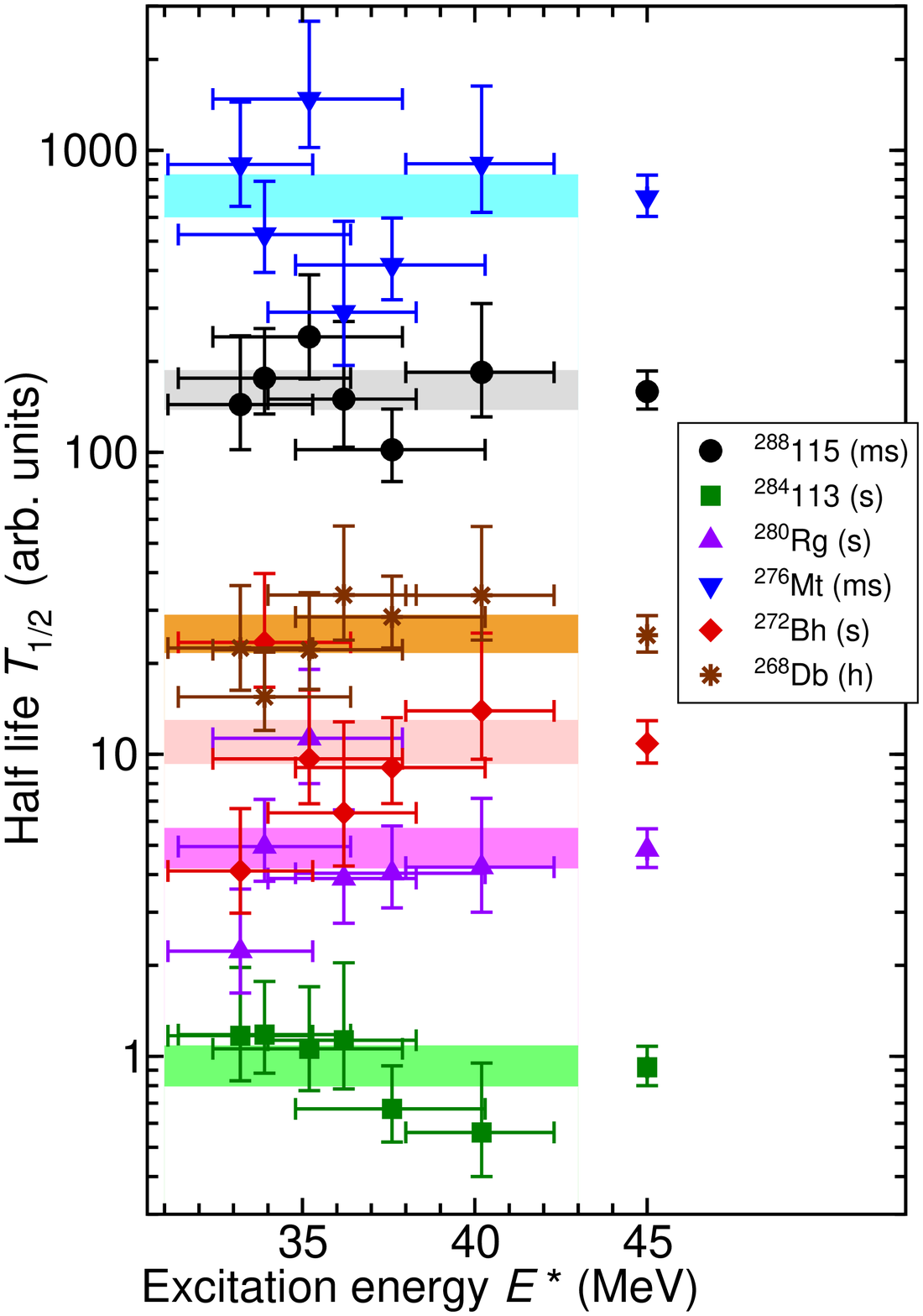}
\caption{
\label{fig4-RC}
(Color online) Half-life analysis for 53 decay chains associated with the production
of $^{288}$115 \protect\cite{rud13,oga13}. The data are split according to 
six different excitation energies, $E^*$, of the compound
system $^{291}$115 \protect\cite{MyS96}. The average is plotted on the right hand side, 
with its error margin represented by the shaded rectangles. See Table SM IV
for details.}
\end{figure}

This Supplemental Material contains one figure and six tables
to support the arguments in the main article. 

Figure~SM~4 and Table~SM~IV provide a consistency check \cite{sch00} 
of the 56 five-$\alpha$ long decay chains attributed to 
the decay of the $3n$ (53 chains) and $4n$ (3 chains) evaporation 
channel, $^{288}$115 and $^{287}$115, following the fusion-evaporation 
reaction $^{48}$Ca+$^{243}$Am at different beam energies
and experiments \cite{rud13,oga13}.

Table~SM~V is an extended version of Table II of the main article.
It includes the results of the test proposed in Ref.~\cite{sch00}
for different decay scenarios and subsets of decay chains associated 
with the eleven observed recoil-$\alpha$-($\alpha$)-chains,
which are considered to stem from element $Z=115$.

Tables~SM~VI-X provide the individual number of a probability check
of the first three decay steps of all hitherto published 67 decay
chains associated with element $Z=115$; Table~SM~VI lists
the complete comparison with respect to the anticipated reference of
the 53 five-$\alpha$ long decay chains originating from $^{288}$115,
while Tables~SM~VII-X relate to the different decay scenarios proposed
and discussed in the main article.

\begin{table*}
\tabcolsep6pt
\caption{
\label{tab:split}
Overview of analyses according to Ref.~\protect\cite{sch00} of 53 five-$\alpha$ chains 
associated with $^{288}$115 at different beam energies \protect\cite{oga13,rud13}. 
The last column summarizes additional three chains associated with $^{287}$115. 
See Fig.~SM~4 for an illustration.}
\begin{tabular}{l|cccccc|c|c}
\hline
&\\[-8pt]
$\langle E_{\rm lab}\rangle$ (MeV)    & 239.8   & 240.8   & 242.1   & 243.4   & 245.0   & 248.1   & ALL & 253.4 \\
$E^*$  \protect\cite{MyS96} &31.1-35.3&31.4-36.4&32.4-37.9&34.0-38.3&34.8-40.3&38.0-42.3&&42.4-47.2\\ 
$d_{\rm target}$ (mg/cm$^2$)& 0.37    &0.84;0.68& 0.83(1) & 0.37    & 0.83(1) &0.36;0.37&&0.36;0.68\\
integral $(10^{18})$        & 11.7    & 10.4    & 2.13(12)& 3.3     &3.89(23) &4.3+3.7  &39.4& 4.3+4.4 \\  
No.~of chains               &  7      & 12      & 8       & 6       & 14      & 3+3     & 53& 1+1(+1)$^a$ \\
$\sigma_{\rm prod}$ (pb)&\multicolumn{2}{c}{$3.5(^{27}_{15})$}&$7.5(10)(^{36}_{26})$&$8.5(^{64}_{37})$&$7.2(9)(^{24}_{19})$& $\sim 4$  &   & $\sim 1$     \\
\hline
&\\[-8pt]
$T_{1/2} (^{288}115)$ (s)   &$0.14(^{10}_4)$&$0.18(^8_4)$&$0.24(^{15}_7)$&$0.15(^{12}_5)$&$0.10(^4_2)$&$0.18(^{13}_5)$&$0.16(^3_2)$&$0.042(^{57}_{15})$\\
data points; $\sigma_{\Theta_{\rm exp}}$      
                            &6 ; 1.70       & 10 ; 1.20  & 7 ; 0.72      & 5 ; 0.98& 14 ; 0.72 {\bf L}& 6 ; 1.20      & 47 ; 1.21  & 3 ; 0.18 {\bf L}  \\
$[\sigma_{\Theta,\rm low},\sigma_{\Theta,\rm high}]$ \protect\cite{sch00} 
                            &[0.48,1.89]    & [0.65,1.82]& [0.52,1.87]   & [0.41,1.90]   & [0.73,1.77]& [0.48,1.89]   & [0.97,1.59]& [0.19,1.91]       \\
\hline
&\\[-8pt]
$T_{1/2} (^{284}113)$ (s)   &$1.17(^{80}_{34})$&$1.18(^{59}_{30})$&$1.06(^{64}_{29})$&$1.13(^{91}_{35})$&$0.67(^{26}_{14})$&$0.56(^{39}_{16})$&$0.92(^{16}_{12})$&$0.082(^{199}_{34})$\\
data points; $\sigma_{\Theta_{\rm exp}}$      
                            &6 ; 1.93 {\bf H}& 9 ; 0.78  & 7 ; 0.78      & 5 ; 1.56      & 13 ; 0.99  & 6 ; 0.53      & 46 ; 1.17  & 2 ; 0.38          \\
$[\sigma_{\Theta,\rm low},\sigma_{\Theta,\rm high}]$ \protect\cite{sch00} 
                            &[0.48,1.89]    & [0.62,1.84]& [0.52,1.87]   & [0.41,1.90]   & [0.72,1.77]& [0.48,1.89]   & [0.96,1.60]& [0.04,1.83]       \\
\hline
&\\[-8pt]
$T_{1/2} (^{280}$Rg) (s)    &$2.2(^{14}_6)$ &$4.9(^{21}_{11})$&$11.3(^{78}_{33})$&$3.9(^{27}_{11})$&$4.0(^{17}_9)$&$4.2(^{29}_{12})$&$4.8(^8_6)$&$0.093(^{223}_{38})$\\
data points; $\sigma_{\Theta_{\rm exp}}$      
                            &7 ; 0.88       & 11 ; 1.11  &6 ; 0.40 {\bf L}&6 ; 0.94      & 11 ; 1.04  & 6 ; 0.84      & 47 ; 1.07  & 2 ; 1.36          \\
$[\sigma_{\Theta,\rm low},\sigma_{\Theta,\rm high}]$ \protect\cite{sch00} 
                            &[0.52,1.87]    & [0.67,1.81]& [0.48,1.89]   & [0.48,1.89]   & [0.67,1.81]& [0.48,1.89]   & [0.97,1.59]& [0.04,1.83]       \\
\hline
&\\[-8pt]
$T_{1/2} (^{276}$Mt) (s)    &$0.90(^{55}_{25})$&$0.53(^{26}_{13})$&$1.48(^{120}_{46})$&$0.29(^{29}_{10})$&$0.42(^{18}_{10})$&$0.90(^{73}_{28})$&$0.70(^{13}_9)$&$0.021(^{28}_8)$\\
data points; $\sigma_{\Theta_{\rm exp}}$      
                            &7 ; 1.25       &  9 ; 1.14  &5 ; 1.62       & 4 ; 1.21      &11 ; 0.62 {\bf L} &5 ; 0.88 & 41 ; 1.15  & 3 ; 0.54          \\
$[\sigma_{\Theta,\rm low},\sigma_{\Theta,\rm high}]$ \protect\cite{sch00} 
                            &[0.52,1.87]    & [0.62,1.84]& [0.41,1.90]   & [0.31,1.92]   & [0.67,1.81]& [0.41,1.90]   & [0.94,1.62]& [0.19,1.91]       \\
\hline
&\\[-8pt]
$T_{1/2} (^{272}$Bh) (s)    &$4.1(^{25}_{11})$&$23.5(^{162}_{68})$&$9.7(^{67}_{28})$&$6.4(^{64}_{21})$&$9.0(^{42}_{22})$&$13.9(^{113}_{43})$&$10.9(^{21}_{15})$&$1.8(^{43}_7)$\\
data points; $\sigma_{\Theta_{\rm exp}}$      
                            &7 ; 1.51       &  6 ; 1.30  & 6 ; 0.76      & 4 ; 1.59      &10 ; 1.12   & 5 ; 0.86      & 39 ; 1.38  & 2 ; 0.19          \\
$[\sigma_{\Theta,\rm low},\sigma_{\Theta,\rm high}]$ \protect\cite{sch00} 
                            &[0.52,1.87]    & [0.48,1.89]& [0.48,1.89]   & [0.31,1.92]   & [0.65,1.82]& [0.41,1.90]   & [0.94,1.62]& [0.04,1.83]       \\
\hline
&\\[-8pt]
$T_{1/2} (^{268}$Db) (h)    &$23(^{14}_6)$&$15(^6_3)$&$22(^{12}_6)$&$34(^{23}_{10})$&$28(^{10}_6)$&$34(^{23}_{10})$&$25(^4_3)$& $1.4(^{18}_5)$\\
data points; $\sigma_{\Theta_{\rm exp}}$      
                            &7 ; 0.77       & 12 ; 0.83  &8 ; 1.04       & 6 ; 0.89      &14 ; 1.20   & 6 ; 0.73      & 53 ; 1.00  & 3 ; 0.79          \\
$[\sigma_{\Theta,\rm low},\sigma_{\Theta,\rm high}]$ \protect\cite{sch00} 
                            &[0.52,1.87]    & [0.70,1.79]& [0.58,1.85]   & [0.48,1.89]   & [0.73,1.77]& [0.48,1.89]   & [0.98,1.58]& [0.19,1.91]       \\
\hline
\end{tabular}

\footnotemark[1]{TASISpec chain measured at 245.0 MeV \protect\cite{rud13} included.}\\
\end{table*}

\begin{table*}
\tabcolsep7pt
\caption{
\label{tab:full}
Half-lives derived from the correlation times of decays of isotopes 
of Rg, $Z=113$ and $Z=115$. Results and confidence intervals of a statistical 
test proposed in Ref.~\protect\cite{sch00} are provided for each half-life 
analysis. The columns labeled `$4n$ channel' and
`$3n$ channel' relate to the three and 53 decay chains associated with these
reaction channels in Refs.~\protect\cite{oga13,rud13,rud14a}, respectively.
The other columns describe different combinations of the decay data from 
recoil-$\alpha$-SF and recoil-$\alpha$-$\alpha$-SF events 
detailed in Table~I. This is an extended version of Table~II.}
\begin{tabular}{cccccccc}
\hline
& \\[-8pt]
data     & $4n$ & $3n$ 
& C1-C7 & C1-C7  & C1-C3,C5-C7 & C4,D4 & C1,C2 \\
selection&\protect\cite{oga13,rud13}&\protect\cite{oga13,rud13}
& D1-D4 &D1,D2,D4& D1,D2       &       &       \\
\hline
& \\[-8pt]
$T_{1/2}(Z=115)$ (s) &
$0.042(^{57}_{15})$ &$0.16(^3_2)$      & $0.39(^{17}_9)$ & $0.26(^{12}_6)$ & 
$0.19(^{11}_5)$  & $0.54(^{129}_{22})$ & $0.11(^{26}_4)$\\
data points; $\sigma_{\Theta_{\rm exp}}$ & 
3 ; 0.18 {\bf L}&47 ; 1.21  & 11 ; 1.19  & 10 ; 1.02   & 8 ; 0.78  & 2 ; 1.65  & 2 ; 0.63\\
$[\sigma_{\Theta,\rm low},\sigma_{\Theta,\rm high}]$ \protect\cite{sch00} &
[0.19,1.91]     &[0.97,1.59]& [0.67,1.81]& [0.65,1.82] &[0.58,1.85]&[0.04,1.83]&[0.04,1.83]\\
\hline
& \\[-8pt]
$T_{1/2}(Z=113)$ (s) &
$0.082(^{199}_{34})$&$0.92(^{16}_{12})$& $2.0(^9_5)$ & $0.63(^{29}_{15})$ &
$0.75(^{41}_{20})$  &$0.17(^{42}_{7})$& $0.35(^{85}_{15})$\\
data points; $\sigma_{\Theta_{\rm exp}}$ &
2 ; 0.38    & 46 ; 1.17   & 11 ; 1.57   & 10 ; 1.21   & 8 ; 0.69  & 2 ; 1.44  & 2 ; 0.02 {\bf L}\\
$[\sigma_{\Theta,\rm low},\sigma_{\Theta,\rm high}]$ \protect\cite{sch00} &
[0.04,1.83] & [0.96,1.60] & [0.67,1.81] & [0.65,1.82] &[0.58,1.85]&[0.04,1.83]&[0.04,1.83]\\
\hline
& \\[-8pt]
$T_{1/2}$(Rg)   (s)  &
$0.093(^{223}_{38})$& $4.8(^8_6)$      & $7.0(^{35}_{18})$ & $2.7(^{15}_7)$ &
 $3.5(^{24}_{10})$  & $0.19(^{46}_8)$  &n/a \\  
data points; $\sigma_{\Theta_{\rm exp}}$ &
2 ; 1.36    & 47 ; 1.07  0 &9 ; 1.90 {\bf H} & 8 ; 1.59& 6 ; 1.18  & 2 ; 0.78  \\
$[\sigma_{\Theta,\rm low},\sigma_{\Theta,\rm high}]$ \protect\cite{sch00} &
[0.04,1.83] & [0.97,1.59] & [0.62,1.84] & [0.58,1.85] &[0.48,1.89]&[0.04,1.83]\\
\hline
\end{tabular}
\end{table*}

\begin{table}
\tabcolsep6pt
\caption{
\label{tab:allvs53}
Practical probability check of the first three decay steps of all 
hitherto published 67 decay chains (Refs.~\protect\cite{rud13,oga13} and 
present data set) associated with the direct production of an isotope of 
element $Z=115$. For each of the decay steps, $i=1,2,3$, $P_i$ is the 
reference probability density function for correlation times on a 
logarithmic scale (see Ref.~\cite{sch00}) evaluated for the measured 
correlation time. The values are normalised with the constant $e$, and 
given in percent. To account for the uncertainties in the reference half-lives,
$P_i$ is given the value 100~\% whenever the correlation time is within
the confidence limits. If it is above (below) the confidence interval, 
the upper (lower) limit is used. 
$\langle P_t\rangle =^n$\hspace*{-1mm}$\sqrt{\prod P_i}$ 
is the geometric average. Reference values 
are $T_{1/2}(Z=115)=0.16(^{3}_{2})$~s, 
$T_{1/2}(Z=113)=0.92(^{16}_{12})$~s, and $T_{1/2}($Rg$)=4.8(^{8}_{6})$~s, 
corresponding to the 
53 five-$\alpha$-long decay chains associated with the decay of $^{288}$115
(cf.~Ref.~\cite{rud14a} and Fig.~1). 
Decay energies are marked '$+$', 'L', 'H', if the measured decay energy is 
compatible with the range $E_1=[10.3,10.6]$~MeV and $E_2=[9.9,10.1]$~MeV or
either too low or too high, respectively. 
These energy ranges are defined either by full-energy 
measurements given in Ref.~\cite{oga13}, and full- or reconstructed energy 
measurements provided in Ref.~\cite{rud13}, cross-checked with Geant4 
simulations \cite{rud13,rud14b,sar14}. 'n/a' denotes incomplete or missing 
data.
}
\begin{tabular}{c|cccc|cc}
\hline
&\\[-8pt]
chain& $P_1$ & $P_2$ & $P_3$ & $\langle P_t\rangle$ & $E_1$ & $E_2$ \\
 ID  & (\%)  & (\%)  & (\%)  & (\%) \\
\hline
\multicolumn{7}{l}{Chains attributed to the $3n$ channel 
in Ref.~\protect\cite{rud13}}\\
  1   & 95 & 98 & 95 & 96 &                  $+$ &  L \\
  2   & 99 & 100 & 98 & 99 &                 $+$ & $+$ \\
  3$^a$& n/a & n/a & 65 & 65 &                 n/a & $+$ \\
  4   & 100 & 100 & 34 & 70 &                $+$ & $+$ \\
  5   & 96 & 46 & 94 & 75 &                  $+$ & $+$ \\
  6   & 60 & 99 & n/a & 77 &                 n/a & $+$ \\
  7   & 64 & 69 & 100 & 76 &                 $+$ & $+$ \\
  8   & 74 & 90 & 60 & 74 &                  n/a & $+$ \\
  9   & 96 & 97 & n/a & 97 &                 $+$ & $+$ \\
 10   & 67 & 74 & 54 & 65 &                  n/a & $+$ \\
 11   & 45 & 22 & 100 & 46 &                 $+$ & $+$ \\
 12   & 100 & 89 & 82 & 90 &                  $+$ & $+$ \\
 13$^a$& 61 & 58 & 23 & 43 &                  n/a & $+$ \\
 14   & 100 & 58 & n/a & 76 &                $+$ & $+$ \\
 15   & 88 & 99 & 94 & 93 &                  $+$ & $+$ \\
 16   & 97 & 36 & 93 & 69 &                  $+$ &  L  \\
 17   & 63 & 90 & 93 & 81 &                  n/a & $+$ \\
 18   & n/a & $<$98 & 100 & $<$99 &                 n/a &  L  \\
 19   & 44 & 43 & 84 & 54 &                  $+$ & $+$ \\
 20   & 61 & 100 & 85 & 80 &                 $+$ & $+$ \\
 21   & 87 & n/a & 99 & 93 &                 $+$ & n/a \\
 22   & 100 & 99 & 30 & 67 &                  n/a & $+$ \\
\hline
\multicolumn{7}{l}{Chains attributed to the $3n$ channel  
in Ref.~\protect\cite{oga13}}\\
  1  & n/a & $<$96 & 24 & $<$48 &                 n/a & $+$ \\
  2  & 69 & 90 & 30 & 57 &                  $+$ & $+$ \\
  3$^a$  & 9.0 & 86 & 56 & 35 &                 $+$ & $+$ \\
  4  & 41 & 100 & 100 & 74 &                $+$ & $+$ \\
  5$^a$  & 35 & 5.0 & 85 & 25 &                 $+$ & $+$ \\
  6$^a$  & 20 & 20 & 99 & 34 &                  $+$ & $+$ \\
  7$^a$  & 9.4 & 62 & 96 & 38 &   $+$ & $+$ \\
  8  & 66 & 65 & 47 & 58 &                  $+$ & $+$ \\
  9$^a$  & 46 & 100 & 19 & 44 &                 $+$ & $+$ \\
 10  & 38 & 98 & 96 & 71 &                  $+$ & $+$ \\
 11  & n/a & $<$100 & 100 & $<$100 &              n/a & $+$ \\
 12  & n/a & $<$100 & 100 & $<$100  &             n/a & $+$ \\
\hline
\end{tabular}
\end{table}
\addtocounter{table}{-1}
\begin{table}
\tabcolsep6pt
\caption{Continued.}
\begin{tabular}{c|cccc|cc}
\hline
&\\[-8pt]
chain& $P_1$ & $P_2$ & $P_3$ & $\langle P_t\rangle$ & $E_1$ & $E_2$ \\
 ID  & (\%)  & (\%)  & (\%)  & (\%) \\
\hline
\multicolumn{7}{l}{Chains attributed to the $3n$ channel 
in Ref.~\protect\cite{oga13}}\\
 13  & 93 & 99 & 95 & 96 &                  $+$ & $+$ \\
 14$^a$  & 18 & 73 & 100 & 50 &                 $+$ &  L  \\
 15$^a$  & 53 & 35 & 83 & 54 &                  $+$ & $+$ \\
 16  & 96 & 98 & 35 & 69 &                  $+$ & $+$ \\
 17  & 86 & 82 & 66 & 77 &                  $+$ & $+$ \\
 18  & 96 & n/a & $<$100 & $<$98 &                $+$ & n/a \\
 19  & 100 & 89 & 96 & 95 &                 $+$ & $+$ \\
 20  & 58 & 93 & 90 & 79 &                  $+$ & $+$ \\
 21$^a$  & 64 & 10 & 100 & 40 &                 $+$ & $+$ \\
 22  & 42 & 100 & 80 & 70 &                 $+$ & $+$ \\
 23$^b$  & 100 & n/a & 28 & 52 &                $+$ & $+$ \\
 24$^a$  & 89 & 32 & 36 & 47 &                  $+$ & $+$ \\
 25  & n/a & $<$67 & 57 & $<$62 &                 n/a & $+$ \\
 26  & 80 & 97 & 75 & 83 &                  $+$ & $+$ \\
 27  & 100 & 100 & 86 & 95 &                $+$ & $+$ \\
 28  & 35 & 63 & 60 & 51 &                  $+$ & $+$ \\
 29  & 73 & 64 & 84 & 73 &                  $+$ & $+$ \\
 30  & 23 & 100 & 96 & 60 &                 $+$ & $+$ \\
 31  & 100 & 78 & 60 & 77   &               $+$ & $+$ \\
\hline
\multicolumn{7}{l}{Chains attributed to the $4n$ channel 
in Refs.~\protect\cite{rud13,oga13}}\\
  1  & 65 & 15 & 0.72 & 8.9 &    $+$ &  H  \\ 
  2  & 49 & n/a & $<$5.4 & $<$16  &               $+$ & n/a \\
  3  & 50 & 31 & 11 & 25               &    $+$ & $+$ \\
\hline
\multicolumn{7}{l}{Recoil-$\alpha$-($\alpha$)-SF chains, 
present data and Ref.~\protect\cite{oga13}}\\
 C1   & 100 & 64 & n/a & 80 &                 $+$ & n/a \\
 C2   & 63 & 63 & n/a & 63 &                  n/a & n/a \\
 C3   & 100 & 100 & 15 & 53 &                 $+$ & $+$ \\
 C4   & 7.0 & 6.0 & 18 & 9.2 &                $+$ & $+$ \\
 C5   & 97 & 63 & 82 & 79 &                   $+$ & $+$ \\
 C6   & 100 & 100 & 100 & 100 &               $+$ & $+$ \\
 C7   & 41 & 91 & 80 & 67 &                   n/a & n/a \\
 D1   & 100 & 100 & 64 & 86 &                 $+$ & $+$ \\
 D2   & 64 & 100 & 72 & 77 &                  $+$ & $+$ \\
 D3   & 0.44 & 0.0020 & 1.2 & 0.10 &           $+$ &  L  \\
 D4   & 55 & 73 & 4.0 & 25  &                 $+$ &  H  \\
\hline
\end{tabular}
\footnotemark[1]{Chain assignment relies also on decay energies and
correlation times of subsequent decay steps \protect\cite{for14}.}\\
\footnotemark[2]{The long-lived $\alpha$ decay assigned to $^{276}$Mt 
in Ref.~\protect\cite{oga13} is associated with $^{280}$Rg 
\protect\cite{rud13}.}
\end{table}

\begin{table}
\tabcolsep6pt
\caption{
\label{tab:shvssc1}
Similar to Table~SM~VI but using the average of the
eleven recoil-$\alpha$(-$\alpha$)-SF chains 
C1-C7 and D1-D4 as reference. 
Reference values are $T_{1/2}(Z=115)=0.39(^{17}_{9})$~s, 
$T_{1/2}(Z=113)=2.0(^{9}_{5})$~s, and $T_{1/2}($Rg$)=7.0(^{35}_{18})$~s
(cf.~column 2 in Table II or column 3 in Table V). Energy ranges are
$E_1=[10.3,10.6]$~MeV and $E_2=[9.5,10.2]$~MeV.
}
\begin{tabular}{c|cccc|cc}
\hline
&\\[-8pt]
chain& $P_1$ & $P_2$ & $P_3$ & $\langle P_t\rangle$ & $E_1$ & $E_2$ \\
 ID  & (\%)  & (\%)  & (\%)  & (\%) \\
\hline
\multicolumn{7}{l}{Recoil-$\alpha$-SF chains, present data and 
Ref.~\protect\cite{oga13}}\\
 C1  & 84 & 40 & n/a & 58 &                      $+$ & n/a \\
 C2  & 35 & 39 & n/a & 37 &                      n/a & n/a \\
 C3  & 90 & 85 & 12 & 45 &                       $+$ & $+$ \\
 C4  & 81 & 3.3 & 15 & 16 &                      $+$ & $+$ \\
 C5  & 98 & 39 & 100 & 73 &                      $+$ & $+$ \\
 C6  & 81 & 81 & 100 & 87 &                      $+$ & $+$ \\
 C7  & 100 & 100 & 71 & 89 &                     n/a & n/a \\
 D1  & 89 & 92 & 55 & 77 &                       $+$ & $+$ \\
 D2  & 36 & 95 & 63 & 60 &                       $+$ & $+$ \\
 D3  & 43 & 6.6 & 20 & 18 &                      $+$ & $+$ \\
 D4  & 30 & 47 & 3.3 & 17  &                     $+$ & $+$ \\
\hline
$\langle$FoM$\rangle$
     &       &       &       & {\bf 52} \\
\hline
\end{tabular}
\end{table}

\begin{table}
\tabcolsep6pt
\caption{
\label{tab:shvssc0}
Similar to Table SM~VII
but isolating chain D3. 
This corresponds to 'scenario 1' illustrated in Fig.~2(a). 
Reference values are $T_{1/2}(Z=115)=0.26(^{12}_{6})$~s,
 $T_{1/2}(Z=113)=0.63(^{29}_{15})$~s, and $T_{1/2}($Rg$)=2.7(^{15}_{7})$~s
(cf.~column 3 in Table II or column 5 in Table V). Energy ranges are
 $E_1=[10.3,10.6]$~MeV and $E_2=[9.9,10.2]$~MeV.
}
\begin{tabular}{c|cccc|cc}
\hline
&\\[-8pt]
chain& $P_1$ & $P_2$ & $P_3$ & $\langle P_t\rangle$ & $E_1$ & $E_2$ \\
 ID  & (\%)  & (\%)  & (\%)  & (\%) \\
\hline
\multicolumn{7}{l}{Recoil-$\alpha$-($\alpha$)-SF chains, present data and 
Ref.~\protect\cite{oga13}}\\
 C1  & 97 & 86 & n/a & 91 &     $+$ & n/a \\
 C2  & 49 & 85 & n/a & 64 &     n/a & n/a \\
 C3  & 100 & 100 & 12 & 49 &                     $+$ & $+$ \\
 C4  & 50 & 9.9 & 15 & 19 &    $+$ & $+$ \\
 C5  & 100 & 85 & 60 & 80 &                      $+$ & $+$ \\
 C6  & 96 & 100 & 95 & 97 &                      $+$ & $+$ \\
 C7  & 91 & 82 & 71 & 81 &                       n/a & n/a \\
 D1  & 99 & 100 & 55 & 82 &                      $+$ & $+$ \\
 D2  & 50 & 99 & 63 & 67 &                       $+$ & $+$ \\
 D4  & 42 & 93 & 3.3 & 23  &   $+$ & $+$ \\
\hline
$\langle$FoM$\rangle$
     &       &       &       & {\bf 65} \\
\hline
 D3  & 16 & 0.0002 & 0.13 & 0.074  &  $+$ & L \\
\hline
\end{tabular}
\end{table}

\begin{table}
\tabcolsep6pt
\caption{
Similar to Table~SM~VII
but assuming that 
ten recoil-$\alpha$-($\alpha$)-SF chains (not D3) have the same origin as
the 53 five-$\alpha$-long chains \protect\cite{rud13,oga13}. 
This corresponds to 'scenario 2' illustrated in Fig.~2(b). 
Reference values are $T_{1/2}(Z=115)=0.18(^{5}_{2})$~s, $T_{1/2}(Z=113)=0.87(^{13}_{10})$~s, 
and $T_{1/2}($Rg$)=4.5(^{7}_{5})$~s [cf.~Fig.~2(b)].
Energy ranges are $E_1=[10.3,10.6]$~MeV and $E_2=[9.9,10.1]$~MeV.
\label{tab:shvssc2}
}
\begin{tabular}{c|cccc|cc}
\hline
&\\[-8pt]
chain& $P_1$ & $P_2$ & $P_3$ & $\langle P_t\rangle$ & $E_1$ & $E_2$ \\
 ID  & (\%)  & (\%)  & (\%)  & (\%) \\
\hline
\multicolumn{7}{l}{Recoil-$\alpha$-($\alpha$)-SF chains, present data and 
Ref.~\protect\cite{oga13}}\\
 C1  & 100 & 66 & n/a & 81 &                  $+$ & n/a \\
 C2  & 57 & 64 & n/a & 61 &                   n/a & n/a \\
 C3  & 100 & 100 & 15 & 53 &                  $+$ & $+$ \\
 C4  & 11 & 6.3 & 19 & 11 &                   $+$ & $+$ \\
 C5  & 99 & 65 & 77 & 79 &                    $+$ & $+$ \\
 C6  & 100 & 100 & 100 & 100 &                $+$ & $+$ \\
 C7  & 50 & 87 & 83 & 71 &                    n/a & n/a \\
 D1  & 100 & 100 & 66 & 87 &                  $+$ & $+$ \\
 D2  & 58 & 100 & 74 & 76 &                   $+$ & $+$ \\
 D4  & 50 & 75 & 4.2 & 25 &                   $+$ &  H  \\
\hline
$\langle$FoM$\rangle$
     &       &       &       & {\bf 64} \\
\hline
 D3  & 0.90 & 0.0007 & 0.72 & 0.076 &  $+$ & L \\
\hline
\end{tabular}
\end{table}

\begin{table}
\tabcolsep6pt
\caption{
Similar to Table~SM~IX
but associating
chains C4 and D4 with a separate origin of decay.
This corresponds to the (preferred) 'scenario 3' illustrated in Fig.~2(c). 
Reference values for these two chains C4 and D4 are 
$T_{1/2}(Z=115)=0.54(^{129}_{22})$~s, $T_{1/2}(Z=113)=0.17(^{42}_{7})$~s, $T_{1/2}($Rg$)=0.19(^{46}_{8})$~s
(cf.~column 5 in Table II or column 6 in Table V),  
$E_1=[10.3,10.6]$~MeV, and $E_2=[9.9,10.2]$~MeV. Reference values for the
remaining eight recoil-$\alpha$-($\alpha$)-SF chains are 
$T_{1/2}(Z=115)=0.16(^{3}_{2})$~s, $T_{1/2}(Z=113)=0.90(^{14}_{11})$~s, $T_{1/2}($Rg$)=4.7(^{7}_{6})$~s
[cf.~Fig.~2(c)],  
$E_1=[10.3,10.6]$~MeV, and $E_2=[9.9,10.1]$~MeV by attributing these
to the same decay origin as 53 five-$\alpha$-long chains 
\protect\cite{rud13,oga13}.
\label{tab:shvssc3}
}
\begin{tabular}{c|cccc|cc}
\hline
&\\[-8pt]
chain& $P_1$ & $P_2$ & $P_3$ & $\langle P_t\rangle$ & $E_1$ & $E_2$ \\
 ID  & (\%)  & (\%)  & (\%)  & (\%) \\
\hline
\multicolumn{7}{l}{Recoil-$\alpha$-($\alpha$)-SF chains, present data and 
Ref.~\protect\cite{oga13}}\\
 C1  & 100 & 65 & n/a & 80 &    $+$ & n/a \\
 C2  & 63 & 63 & n/a & 63 &     n/a & n/a \\
 C3  & 100 & 100 & 15 & 53 &                     $+$ & $+$ \\
 C5  & 97 & 64 & 79 & 79 &                       $+$ & $+$ \\
 C6  & 100 & 100 & 100 & 100 &                   $+$ & $+$ \\
 C7  & 41 & 89 & 81 & 67 &                       n/a & n/a \\
 D1  & 100 & 100 & 65 & 87 &                     $+$ & $+$ \\
 D2  & 64 & 100 & 73 & 78                     &  $+$ & $+$ \\
\hline
C4 & 100 & 41 & 100 & 74 & $+$ & $+$ \\
D4 & 28 & 100 & 88 & 63 &  $+$ & $+$ \\
\hline
$\langle$FoM$\rangle$
     &       &       &       & {\bf 74} \\
\hline
\end{tabular}
\end{table}